\documentclass{elsart5p}

\usepackage{graphics}
\usepackage{graphicx}
\usepackage{amssymb}
\usepackage{amssymb}
\usepackage{amsmath}
\newcommand{\bra}[1]{\langle #1|}
\newcommand{\ket}[1]{|#1\rangle}

\begin{document}

\begin{frontmatter}



\title{Berry Phases, Quantum Phase Transitions and Chern Numbers}
%
\author[AA]{H.A. Contreras},
\author[AA]{A.F. Reyes-Lega\corauthref{Name1}}
\ead{anreyes@uniandes.edu.co}

\address[AA]{Departamento de F\'isica, Universidad de los Andes\\
Cra. 1E No. 18A-10. Edificio H. A.A. 4976\\
 Bogot\'a D.C. (Colombia)}

\corauth[Name1]{Corresponding author. Tel: (571) 332-4500 fax: (571)
332-4516}

\begin{abstract}
We study the relation between Chern numbers  and Quantum Phase
Transitions (QPT) in the XY spin-chain model.  By coupling the spin
chain to a single spin, it is possible to study topological
invariants associated to the coupling Hamiltonian. These invariants
contain global information, in addition to the usual one (obtained
by integrating the Berry connection around a closed loop). We
compute these invariants (Chern numbers) and discuss their relation
to QPT. In particular we show that  Chern numbers can be used to
label regions corresponding to different phases.
\end{abstract}


\end{frontmatter}
Recently, a close connection between Berry phases (BP) associated to
quantum many-body systems and Quantum Phase Transitions (QPT) has
emerged, attracting much attention
\cite{carollo-pachos,zhu,dicke,hamma,Loschmidt}. From an
experimental point of view such a connection is very interesting,
due to the robustness of BP against continuous changes in the
system's parameters. In particular, it has been proposed by Carollo
et al. \cite{carollo-pachos} that the existence of a QPT could be
detected, without having to undergo a phase transition, through a
cyclic evolution in parameter space. In that work, the authors have
shown that a BP can be defined
 for the XY model whose behavior with respect to the system's
parameters reflects the presence of QPT. They computed the relative
geometric phase between the ground and first excited states for
loops around the XX criticality region and found that its derivative
presents a singular behavior at the critical point. Zhu~\cite{zhu},
working also with the XY model, obtained similar results for the
derivative of the BP corresponding to the ground state. He also
analyzed the scaling behavior of the BP, showing that it can be used
as a signature of quantum criticality. Plastina et al.~\cite{dicke}
found, using the Dicke model, that the BP vanishes exactly (in the
thermodynamic limit) in the region corresponding to the normal phase
and is greater than zero in the super-radiant phase. They also found
a singular behavior of the derivative of the BP at the critical
point. Yet another proposal to detect QPT using Berry phases has
been put forward by Hamma~\cite{hamma}.

All these examples point towards a very general, model independent
relation between BP and QPT. In order to understand the origin of
this relation, it is necessary to recognize general patterns that
may emerge from concrete examples.  It is a well known fact, first
pointed out by Simon~\cite{simon} that given a parameter-dependent
Hamiltonian $H(\alpha)$, the BP corresponding to the $n^{th}$
(non-degenerate) energy band $E_n(\alpha)$, arises as the holonomy
of a connection defined on the line bundle spanned by the family of
all eigenvectors $|\varphi_n(\alpha)\rangle$, as $\alpha$ varies
over the parameter space, where
\begin{equation}
H(\alpha)|\varphi_n(\alpha)\rangle=
E_n(\alpha)|\varphi_n(\alpha)\rangle.
\end{equation}
But in the examples related to QPT studied so far, it is not clear
what -exactly- the base  space of the bundle is. In fact, all
computations rely on the introduction of an additional parameter
through a unitary transformation of the Hamiltonian. Then the Berry
phase for a special class of loops in this extended parameter space
is computed, leading to the results reported in references
\cite{carollo-pachos,zhu,dicke,hamma,Loschmidt}.

In this work, using the model proposed in \cite{Loschmidt},  we
intend to go a step further in the sense that, for each value of the
external field ($\lambda$), we identify a  space (isomorphic to the
two-sphere) and a corresponding line bundle $L_\lambda$ over it, for
which a topological invariant $c_1(\lambda)$ can be computed. As
shown below, this invariant is closely related to the phase
transition of the $XY$ model at the critical value $\lambda=1$.

Following Yuan et al.~\cite{Loschmidt}, we start by considering a
system consisting of a spin chain coupled to a central spin, with
Hamiltonian $H=H_{\mbox{\small I}}+H_{\mbox{\small
C}}+H_{\mbox{\small E}}$, where
\begin{equation}
\label{eq:2} H_{\mbox{\small C}}=\frac{\mu}{2}\sigma^z +
\frac{\nu}{2}\sigma^x,
\end{equation}
\begin{equation}
\label{eq:3}
 H_{\mbox{\small E}}\hspace{-0.1cm}=\hspace{-0.1cm}-J\sum_{j=1}^{N} \left (
\frac{1+\gamma}{2} \sigma_{j}^x \sigma_{j+1}^x + \frac{1-\gamma}{2}
\sigma_{j}^y \sigma_{j+1}^y +\lambda \sigma_{j}^z \right ),
\end{equation}
\begin{equation}
\label{eq:4} H_{\mbox{\small I}}=\frac{J g}{N}\sum_{j=1}^N \sigma^z
\sigma_{i}^z.
\end{equation}
The Pauli matrices $\sigma^\alpha$ and $\sigma_j^\alpha$
($\alpha=x,y,z$) describe, respectively, the central spin and the
spin on the $j^{\mbox{\small th}}$ site of the environmental spin
chain. As can be read from eqns. (\ref{eq:2})-(\ref{eq:4}), the
parameters $J,\lambda, g, \mu,\nu$ describe the coupling to the
external field and the strength of the interaction among spins. The
parameter $\gamma$ accounts for the anisotropy in the spin chain.
Assuming that the spin chain is in its ground state, the following
effective mean-field Hamiltonian for the central spin can be
obtained~\cite{Loschmidt}:
\begin{equation}
H_{\mbox{\small eff}}=\left( \frac{\mu}{2} +
\frac{2Jg}{N}\sum_{k=1}^{N/2}\cos \theta_k\right)\sigma^z +
\frac{\nu}{2}\sigma^x.
\end{equation}
Here $\theta_k$ satisfies $\cos \theta_k
=(J\epsilon_k)/\sqrt{\epsilon_k^2+\gamma^2\sin^2\left(\frac{2\pi
k}{N}\right)}$, with
\begin{equation}
\epsilon_k=\lambda-\cos\left(\frac{2\pi
k}{N}\right)+\frac{g\mu}{N\sqrt{\mu^2+\nu^2}}.
\end{equation}
 Following references
\cite{carollo-pachos,zhu,dicke,hamma,Loschmidt}, we change the
Hamiltonian by means of a unitary transformation
$U(\varphi)=\exp\left(-i\varphi\sigma^z/2\right)$ to
\begin{equation}
\label{eq:1} \widetilde H=U(\varphi)H_{\mbox{\small
eff}}U^{\dagger}(\varphi).
\end{equation}
If we keep all parameters appearing in $\widetilde H$ fixed, with
the exception of $\varphi$ and $\gamma$, we can regard this
Hamiltonian as defined on the surface of a two dimensional sphere,
obtained by stereographic projection from the plane with polar
coordinates $0 \leq\gamma\leq\infty$ and $0\leq \varphi<2\pi$, given
that we restrict the parameter $\nu$ to the limiting case $\nu\ll
1$. The compactification of the $\varphi$-$\gamma$ plane to a sphere
is possible in that limiting case, because then the eigenvectors of
$\widetilde H$ do not depend on $\varphi$ when
$\gamma\rightarrow\infty$. As mentioned above a parameter-dependent
Hamiltonian induces, for each non-degenerate energy band, a line
bundle over the parameter space. This line bundle comes equipped
with a connection whose holonomy is precisely the geometric or Berry
phase corresponding to the given eigenstate~\cite{simon,avron}. In
the present case, we are regarding the Hamiltonian $\widetilde H$ as
depending on the two parameters $\varphi$ and $\gamma$.
 The ground state of $\widetilde
H$ is readily shown~\cite{Loschmidt} to be given by:
\begin{equation}
\ket{g(\gamma,\varphi)}=\left(\sin (\psi/2)e^{ i\varphi},
-\cos(\psi/2)\right),
\end{equation}
where
\begin{equation}
\sin \psi=\nu/\sqrt{\nu^2 +( \mu +
4Jg\mbox{$\sum_{k=1}^{N/2}$}\cos\theta_k/N)^2}
\end{equation}
This ground state generates a line bundle $L_\lambda$ over the
sphere, whose topology is characterized by the first Chern number, a
topological invariant that can be expressed as an integral over the
parameter space (two-sphere) as~\cite{avron}:
\begin{equation}
c_1(\lambda)= \frac{-i}{2\pi }\int P dP \wedge dP,
\end{equation}
where $P$ denotes the projector
$P=\ket{g(\gamma,\varphi)}\bra{g(\gamma,\varphi)}$.

After a  long but straightforward calculation we obtain, in the
thermodynamic limit,
 \begin{equation}\label{eq:5}
c_1(\lambda)=\left\{\begin{array}{cc}
    \frac{1}{2} \left ( \mbox{sign} \left (  \mu-2 J g \right )-\mbox{sign}(\mu) \right ),&   \lambda<-1 \\
   \frac{1}{2} \left (  \mbox{sign} \left ( \chi \right )-\mbox{sign}(\mu) \right ), &\lambda \in [-1,1] \\
  \frac{1}{2} \left (   \mbox{sign}\left ( \mu+2 J g \right )-\mbox{sign}(\mu) \right ), &  \lambda>1,\\
       \end{array}\right.
\end{equation}
with
\begin{equation}
\chi=\mu+\frac{ 4 J g}{\pi} \arcsin(\lambda).
\end{equation}
Equation (\ref{eq:5}) is the main result of this paper. It shows that a
topological invariant can be extracted from the effective
Hamiltonian $\widetilde H$ that contains information about the QPT
of the environmental spin chain at $|\lambda|=1$. The interest of
this result lies in the fact that it may be possible, in a general
case, to express $c_1(\lambda)$ as a function of the expectation
values of certain physical observables. In contrast to BP, that
depends not only on the topology of the state space but on the path
followed in parameter space, the Chern number is a purely
topological quantity, an integer that is robust against continuous
perturbations of the Hamiltonian. This same idea can be applied to
the ground state of the $XY$ spin chain (without the coupling to a
central spin). In that case, the Chern number scales with the number
$N$ of spins, but after normalization with $N$, one obtains a
function that also reflects the presence of a QPT at the critical
point. This result, and potential applications thereof, will be
reported elsewhere.

\section*{Acknowledgement}
The authors gratefully acknowledge discussions with F.J. Rodr\'iguez
and L. Quiroga. Financial support from the Faculty of Sciences of
Universidad de los Andes is acknowledged.


\begin{thebibliography}{99}

\bibitem{carollo-pachos}
A. Carollo et al., \textit{Phys. Rev. Lett.} {\bf 95}, 157203 (2005)
\bibitem{zhu}
S.-L. Zhu, \textit{Phys. Rev. Lett.} {\bf 96}, 077206 (2006)
\bibitem{dicke}
F. Plastina et al., \textit{Europhys. Lett.}, {\bf 76}, 182 (2006)
\bibitem{hamma}
A. Hamma, pre-print: quant-ph/0602091
\bibitem{Loschmidt}
Z.-G. Yuan et al., \textit{Phys. Rev. A} {\bf 75}, 012102 (2007)
\bibitem{simon}
B. Simon, \textit{Phys. Rev. Lett.} {\bf 51}, 2167 (1983)
\bibitem{avron}
J.E. Avron et al., \textit{Comm. Math. Phys.} {\bf 124}, 595 (1989)
\end{thebibliography}
\end{document}